\documentclass[12pt,preprint]{aastex}

\newcommand{\tcool}{t_\mathrm{cool}}
\newcommand{\bcrit}{\beta_\mathrm{crit}}
\newcommand{\qcrit}{Q_\mathrm{crit}}

\def\ee #1 {\times 10^{#1}}          
\def\ut #1 #2 { \, \mathrm{#1}^{#2}} 
\def\u #1 { \, \mathrm{#1}}          
\def\msol{\hbox{$\hbox{M}_\odot$}}
\def\lsol{\hbox{$\hbox{L}_\odot$}}
\def\kms{km s$^{-1}$}
\def\etal   {{et al.}}                     

\begin{document}


\shorttitle{}
\shortauthors{}

\title{On the Origin of the Central 1$''$ Hole in the Stellar Disk of Sgr A*\\
 and the \emph{Fermi} Gamma-Ray Bubbles}
\author{Mark Wardle$^1$ and Farhad Yusef-Zadeh$^2$}
\affil{$^1$Department of Physics \& Astronomy and Research Centre for Astronomy, Astrophysics \\ \& Astrophotonics, Macquarie University, 
Sydney NSW 2109, Australia}
\affil{$^2$Department of Physics \& Astronomy and Center for Interdisciplinary Exploration \\ \& Research in Astrophysics, Northwestern University, Evanston, IL 60208}


\begin{abstract} 
The supermassive black hole Sgr A* at the center of the Galaxy is surrounded by two misaligned disks of young, massive stars extending from $\sim 0.04$ to 0.4\,pc.  The stellar surface density increases as $\sim r^{-2}$ towards Sgr A* but is truncated within 1$''$ (0.04pc).  We explore the origin of this annulus using a model in which star formation occurs in a disk of gas created through the partial capture of a gas cloud as it sweeps through the inner few parsecs of the galaxy and temporarily engulfs Sgr A*.   We identify the locations within which star formation and/or accretion onto Sgr A* take place.   Within 0.04\,pc the disk is magnetically active and the associated heating and enhanced pressure prevents the disk from becoming self gravitating.  Instead, it forms a magneto-turbulent disk that drains onto Sgr A* in $\la 3\times10^{6}$ years.  Meanwhile, fragmentation of the gas beyond the central 0.04\,pc hole creates the observed young stellar disk. The two large scale bubbles of $\gamma$-ray emission extending perpendicular to the Galactic plane may be created by a burst of accretion of $\sim 1\times10^{5}\msol$ of gas lying between 0.01 and 0.03\,pc.   The observed stellar ages imply that this capture event occurred $\sim 10^{6.5}$\,yr ago, thus such events occurring over the life time of the Galaxy could have significantly contributed to the current mass of Sgr A* and to the inner few parsec of the nuclear star cluster.  We suggest that these events also occur in extragalactic systems.
\end{abstract}


\section{Introduction}

Two bubbles of $\gamma$-ray emission with sharp edges extend symmetrically away from the Galactic plane up to Galactic latitudes b$\sim\pm50^0$ (Dobler et al. 2010; Su et al. 2010). This gigantic structure is narrower near the Galactic plane and appears to emanate from  the Galactic center. The $\gamma$-ray luminosity, $\sim4\times10^{37}$ erg\, s$^{-1}$,  requires a total energy input of 10$^{54-55}$ ergs (Su et al. 2010). Two classes of model have been proposed to explain the origin of this emission. In one picture, the Fermi $\gamma$-ray bubbles are a relic of past AGN-like activity stimulated by accretion of stars or gaseous material onto the central supermassive black hole, Sgr A*  (e.g.\ Zubovas, King and Nayakshin 2011; Cheng et al. 2011; Guo et al. 2012; Yang et al. 2012, 2013). In the other, a nuclear starburst drives a powerful wind  from the inner region of the Galaxy that inflates the bubbles (e.g.\ Crocker and Aharonian 2010; Crocker 2012; Carretti et al. 2013).

The emission from Sgr~A* is thought to arise from a radiatively inefficient accretion flow, with a possible contribution from outflows or jets (e.g., Blandford and Begelman 1999; Falcke and Markoff 2000; Yuan, Falcke and Markoff 2002).  However, the possible association of the Fermi gamma-ray bubbles with Sgr A* implies that periods of increased accretion onto Sgr A* may produce powerful outflow and jet driven activity (Zubovas and Nayakshin 2012).  This activity requires high accretion rate and a reservoir of gaseous material feeding Sgr A*. 

Indeed, there is evidence for an accretion event occurring a few million years ago in the form of one or perhaps two counter-rotating disks  of young massive stars orbiting between 0.04 and 0.4\,pc of Sgr A*.  The stellar ages and total mass are estimated to be $\sim 6$\, Myr and  $\sim 1.5\times 10^4\,\msol$, respectively (Paumard et al.\ 2006; Lu et al. 2009; Yelda et al.\ 2014).  The very high density ($\ga 10^{10}\ut cm -3 \; [r(\mathrm{pc})]^{-3}$) required for self-gravity to overcome tidal shear in the vicinity of the black hole implies that the stars were formed by fragmentation of a self-gravitating disk (Levin \& Beloborodov 2003), presumably from a captured molecular cloud (Nayakshin \& Cuadra 2005; Nayakshin et al. 2007; Bonnell \& Rice 2008; Wardle \& Yusef-Zadeh 2008, 2012, hereafter WY08 and WY12; Alig et al. 2011; Mapelli et al. 2012; Lucas et al. 2013).

This suggests a picture in which a byproduct of this process was the accretion of gas by Sgr A*, and that the associated outburst provides the needed energetics to produce the Fermi Bubbles (Zubovas, King \& Nayakshin 2011; Zubovas \& Nayakshin 2012).   These models assume that roughly half of the captured gas is converted to stars with the remainder being accreted onto the black hole.  Alexander et al. (2012) pointed out that the absence of stars within an arc second suggests that the inner section of the captured disk was accreted by Sgr A*.  However, the estimated timescale for accretion far exceeded the $\sim10^6$--$10^7$ year time frame required by the age of the stars and the time scale to power the Fermi bubbles.

Here we re-examine this issue in the context of the compact disks formed by partial cloud capture when an extended cloud temporarily engulfs Sgr A* while on a passage through the central few parsec of the Galaxy.  Previous simple analytic estimates show that  $\sim 10^4$--$10^5\msol$ of gas settles into a sub-parsec disc (WY08, WY12). Here we show that this capture process produces surface density profiles that are steeper than $\Sigma\propto r^{-3/2}$.  Magnetic activity prevents fragmentation within the inner regions, driving accretion at rates $\sim 0.01$\,\msol\,yr$^{-1}$ compatible with the few million-year time scale for formation of the \textit{Fermi} gamma ray bubbles.  Meanwhile the outer disk fragments to form stars (see Fig.\ 1).

\section{Disk Model}

We begin our analysis by estimating the surface density profile of the cloud material as it circularizes, cools and settles into a disk.  We envisage,  following WY08, that a disk of gas surrounding Sgr A* is formed via  partial capture of a gas cloud that sweeps across the black hole with speed $v$.  Cloud material with impact parameter less than the critical value $b_0 = 2GM/v^2$ is captured, where we adopt $M=4\times10^{6}\,\msol$ for the mass of Sgr A*.  Partial Hoyle-Lyttelton-like angular momentum cancellation between material passing on opposite sides of Sgr A* leading to the formation of a sub-parsec disk with mass $\sim 10^4$--$10^5\msol$ (WY08).  During the process of circularization, the components of angular momentum perpendicular to the eventual disk's rotation axis are eliminated.  We assume that on average each fluid element ends up with a fraction $\lambda$ of its original specific angular momentum parallel to the final rotation axis.  Thus a fluid element with initial impact parameter $b$ perpendicular to the eventual rotation axis ends up on a Keplerian orbit with specific angular momentum $\sqrt{GMr} = \lambda b v$.   W08 adopted a nominal value $\lambda_0 = 0.3$, yielding disk masses $M_d=\pi b_0^2 \,\Sigma_\mathrm{cloud}$ and radii $r_d = 2 \lambda_0 \, b_0^2$, that were later found to be consistent with those arising in simulations of cloud capture by this mechanism (Mapelli et al 2012).

Here we generalise the W08 model by setting $\lambda = \lambda_0 \,(b/b_0)^p$ for some $p>0$.  The physical motivation for this prescription is that  cancellation should become less effective for material with larger impact parameters because of the cloud's internal structure. We can then estimate the disk's surface density profile $\Sigma(r)$ by equating the mass of captured gas with impact parameters in the range $[b,b+db]$, i.e.\  $4\,\Sigma_\mathrm{cloud} \sqrt{b_0^2-b^2} \,\, db$, to the mass of disk material at radii in $[r,r+dr]$, i.e.\ $2\pi r \Sigma(r) \,dr$, with the prescription for angular momentum loss yielding $r = (\lambda b v)^2/GM$.   Then 
\begin{equation}
    \Sigma(r) = \frac{qM_d}{\pi^2 r_d^2}\,\left(\frac{r}{r_d}\right)^{\!\frac{1}{2}q-2} 
                   \!\!\sqrt{1-\left(\frac{r}{r_d}\right)^{\!\!q}}\,,
    \label{eqn:Sigma}
\end{equation}
where $q=1/(1+p)$.   We choose $\lambda_0=0.3$ for consistency with the simulated disk masses and radii, and adopt $p=1$, noting that the profile is insensitive to $p$. Note that the mass, radius and profile are broadly consistent with the inferred $\sim r^{-2}$ stellar profile, for reasonable parameters of the incoming cloud.  For example, a cloud column density $1\times10^{24}$ cm$^{-2}$ and velocity 120 \kms\, yields $M_d\approx 2\times10^5$ \msol and $r_d\approx0.4$\,pc.

The fate of the disk is determined by the competition between heating and radiative cooling.  Heating by starlight or dissipation of magnetically-driven turbulence may keep the disk warm enough that self gravity is not important.  Should the temperature drop to the point that Toomre's parameter 
\begin{equation}
    Q = \frac{c_s\Omega}{\pi G \Sigma} < \qcrit\approx 1\,,
    \label{eqn:Q}
\end{equation}
where $\Omega$ is the orbital frequency and $c_s$ is the sound speed, the disk will fragment if it can quickly radiate away the thermal energy liberated during gravitational collapse, i.e.
\begin{equation}
    \Omega \,\tcool < \bcrit\,,
    \label{eqn:beta} 
\end{equation}
where $\tcool$ is the local cooling time scale (Gammie 2001) and we adopt  $\bcrit=3$, as suggested by simulations of self-gravitating disks (see the review by Lodato 2012 and references therein).\footnote{Recent simulations suggest $\bcrit\sim 10$, and it may even be that fragmentation happens stochastically for any value of $\bcrit$  (Pardekooper 2012; Rice et al.\ 2012; Meru \& Bate 2012). Fortunately our results turn out to be insensitive to this uncertainty:  the limited gravitoturbulent region in Fig. 3 shrinks as $\bcrit$ is increased, vanishing when $\bcrit\ga 30$.}  However, if cooling is inefficient  the disk enters a ``gravitoturbulent'' state in which self-gravity continually perturbs the density and velocity field, and heating via dissipation of weak turbulence maintains $Q\approx\qcrit$ (Paczynski 1978; Gammie 2001; Rafikov 2009).
  
We first consider radiative losses, which depend on the optical depth $\kappa$ between the mid plane and surface, $\tau = \kappa\Sigma/2$, which here is much larger than unity.    For $T>150$\,K we adopt the piecewise power-law opacities  provided in Table 1 of Zhu et al.\ (2009).  As these do not include icy grains which dominate below 150\,K at lower temperatures we follow Rafikov (2009) in adopting  $\kappa = 5\times10^{-4}T^2 \ut cm 2 \ut g -1 $ where $T$ is in Kelvin, which approximates the calculations of Semenov et al.\ 2003).
The radiated power per unit area is
\begin{equation}
    F = \frac{4\sigma  T^4}{\kappa \Sigma} \,,
    \label{eqn:F}
\end{equation} 
the sum of $\sigma T^4/\tau$ from each of the upper and lower surfaces.  Cooling in a self-gravitating disk occurs at constant pressure $\pi G\Sigma^2/2$ (Paczynski 1978), yielding a time scale
\begin{equation}
    t_\mathrm{cool} \approx \frac{7}{2}\,\frac{\Sigma c_s^2}{F} \approx  \frac{7}{8}\frac{\kappa c_s^2 \Sigma^2}{\sigma T^4}\,.
    \label{eqn:tcool}
\end{equation}
where we have adopted adiabatic index $\frac{7}{5}$, appropriate for warm molecular gas.  

Turning now to heating processes, accretion at rate $\dot{M}$ yields a dissipative power
\begin{equation}
    D = \frac{3}{4\pi}\dot{M}\Omega^2
    \label{eqn:D}
\end{equation} 
per unit area of the disk (e.g.\ Pringle 1981).  We parametrize the accretion rate in the standard fashion using the $\alpha$ parameter:
\begin{equation}
    \dot{M} = 3\pi\,\alpha \,\Sigma c_s^2/\Omega\,.
    \label{eqn:dotM}
\end{equation}
For gravitoturbulence, $\alpha$ is just that required to maintain $Q=\qcrit$ (Paczynski 1978). Setting $D=F$, and using eqs (\ref{eqn:tcool}),(\ref{eqn:D}),(\ref{eqn:dotM}), we obtain
\begin{equation}
    \alpha = \frac{14}{9}\,\Omega\,\tcool
    \label{eqn:alphaG}
\end{equation}
(cf.\ Gammie 2001).

Alternatively, accretion will be driven by magnetic stresses if the level of ionization is sufficient.  As in protostellar disks (e.g.\ Gammie 1996; Wardle 2007), external ionizing sources such as cosmic rays, X-rays and UV are ineffective, and coupling can only be sustained by thermal ionization of potassium, magnesium and sodium which becomes effective above $\approx 900$\,K (e.g. Umebayashi 1983). When this is the case, shear in the disk maintains a strong azimuthal magnetic field developed from the incoming cloud's pre-existing field, and this efficiently transports angular momentum, yielding $\dot{M} = 3\pi\,\tilde{\alpha} \, h^2\Omega \Sigma$, where $2h$ is the disc thickness and $\tilde{\alpha}\sim 0.1$--$0.2$   (Gaburov at al.\ 2012; Bai \& Stone 2013).  Crucially, the magnetic pressure in the disk is about an order of magnitude larger than the gas pressure, inflating the disk so that $h\approx 3c_s/\Omega$.  In addition, for a given temperature and column density, the volume density and hence the effective value of $\qcrit$ are all reduced by a factor of three.  Thus when the temperature exceeds 900\,K, we set
\begin{equation}
    \alpha =1 \,,\quad\textrm{and}\quad \qcrit=0.3 \,.
    \label{eqn:alphaQ-forB}
\end{equation}

The disk would also have been heated by hot stars lying within a few parsec of Sgr A*, which at present have a net luminosity $\approx 2\times10^7\lsol$ (Krabbe et al 1995; Latvakoski et al 1999).  As many of these stars were presumably created by fragmentation of the disk we adopt half of the present-day value, i.e.
\begin{equation}
    D_* = \frac{10^7 \lsol}{4\pi (1\u pc )^2}
    \label{eqn:Dstar}
\end{equation}
as a rough estimate of the heating rate per unit area, independent of distance from Sgr A*\footnote{Although stellar heating at the time of disk formation may have been even lower, it has little effect on fragmentation within 0.5\,pc of Sgr A* (see the upper panel of Fig.~4).}.

\section{Results}

We first consider how the disk's fate at $r=0.04$\,pc depends on the local column density, with the aim of determining the column that would place the inner boundary of the fragmentation region there (see Fig.\ 2).  If $\Sigma\ga 600$\,g\,cm$^{-2}$, the high optical depth allows magnetic heating to maintain the temperature above 900\,K, and thermal ionization allows continued magnetic activity.  Inflation by magnetic pressure then implies that $\qcrit\approx0.3$ and as this corresponds to temperatures below 600\,K we conclude that self-gravity is unimportant.  On the other hand, when $\Sigma\la 600$\,g\,cm$^{-2}$ magnetic heating is unable to maintain the temperature above 900\,K,  thermal ionization is insufficient to couple the magnetic field to the gas, magnetic activity shuts down, and $\qcrit=1$.   For $\Sigma\la 250$\,g\,cm$^{-2}$ stellar radiation keeps the disk hot enough to avoid becoming self-gravitating, whereas at higher column densities the disk is able to cool to the point that self-gravity is important, and between $\sim 300$ and $600$\,g\,cm$^{-2}$ the disk cools rapidly enough to fragment.

Fig.\ 3 shows how the disk behavior depends on column density and distance from Sgr A*.  In general, high surface densities are unstable to fragmentation, while low surface densities are neither able to fragment nor to accrete because stellar heating maintains $Q\ga 1$ but with $T<900$\,K so that magnetic coupling is ineffective. Within about 0.05\,pc of Sgr A* there is an intermediate range of column densities for which magnetic activity maintains $T>900$\,K and prevents fragmentation.  Note that disks are gravitoturbulent over a severely limited range of radii and surface densities. 

The blue dashed curve shows the surface density profile (eq 3) for a disk mass
$M_d=2\times10^5\,\msol$, corresponding to $\Sigma \approx 600$\,g\,cm$^{-2}$ at 0.04\,pc, so that the inner boundary of the fragmentation region is at 0.04\,pc, consistent with the hole in the stellar distribution around Sgr A*.    In light of the very rapid cooling we expect that the gas will fragment well before settling into the well-ordered disk we implicitly assume here, producing a dynamically warm stellar distribution.   Fragmentation very near the outer edge of the disk may be prevented by stellar heating, although the exact profile of the disk edge and the stellar heating rate are both rather uncertain.  Such a disk would be magnetically active interior to 0.04 pc, with initial accretion rates 0.01--0.03\,\msol\,yr$^{-1}$, as indicated by the red contours in Fig.~3.

Fig.\ 4 shows the radial structure of this disk.  The top panel compares the temperatures maintained by magnetic activity and stellar heating with those at which the disk becomes self-gravitating and the cooling time scale is 3 or 10 $\Omega^{-1}$.  The lower panel of Fig.\ 4 shows the radial profile of surface density and optical depth. The latter is calculated for the equilibrium temperature maintained by magnetic dissipation within 0.04\,pc, or the temperature corresponding to $Q=1$ at larger radii.  The discontinuities in optical depth arise due to evaporation of ice mantles at 150\,K, the sublimation of grains at $\sim 1200$\,K, and  collisional dissociation of water at $\sim 1500$\,K. The fragmentation region contains about $6\times 10^4\,\msol$ of gas, consistent with estimates of the stellar mass in the disk (Lu et al.\ 2013).  The characteristic initial fragment mass is $(2 \pi c_s/\Omega)^2 \, \Sigma \sim2.5$\,\msol, but this is expected to grow by about an order of magnitude by collisions (Levin \& Beloborodov 2003).  Meanwhile the disk within 0.04 \,pc exhibits accretion rates $\sim0.01$--0.03\,\msol\,yr$^{-1}$, with accretion time scales $\sim 1$--3\,Myr that are an order of magnitude shorter than previously estimated (Alexander et al.\ 2012) because of the enhancement by magnetic levitation.

\section{Discussion}

We have examined in more detail the earlier proposal in which gas clouds engulfing supermassive black holes leave behind a captured disk (WY08, WY12). The angular momentum cancellation inherent in this scenario naturally produces steep ($\sim r^{-1.75}$) surface density profiles consistent with the observed stellar disk, with $10^4$--$10^5$\,\msol\ and size $\sim0.5$\,pc (W08).  For reasonable parameters we found that such a captured disk would indeed be unstable to fragmentation between 0.04 and 0.4\,pc, consistent with the sizes of the observed stellar disks.  Between 0.01 and 0.04\,pc dissipation of magnetically-driven turbulence prevents the disk from becoming self-gravitating and enables accretion with $\dot{M} \approx 1$--$3\ee -2 \,\msol\ut yr -1 $.  

The magnetized accretion disk that arises in the central 0.04\,pc explains the central 1$''$ hole in the stellar distribution (cf.\ Alexander et al.\ 2012).  Of significance here is that because magnetic pressure dominates gas pressure (Gaburov et al.\ 2012), thickening the disk and increasing the turbulent velocities by  a factor of three,   the accretion rate is an order of magnitude larger than in ``standard'' accretion disk models (e.g.\ Alexander at al.\ 2012).   This increases the plausibility of accretion models for the origin of the Fermi bubbles, notably by reducing the accretion time scale to a few million years, enabling significant accretion to occur between formation of the stellar population and the Fermi bubbles.   The energy released by such an event is uncertain.  The current accretion rate onto Sgr A* is thought to be $\sim 10^{-5}\,\msol\ut yr -1 $ and the bolometric luminosity $L\sim 150\lsol$ (see Genzel, Eisenhauer \& Gillesen 2010 and references therein), yielding an efficiency $L/\dot{M}c^2 \sim10^{-6}$.  If the accretion of $\sim10^{5}\,\msol$ is to power the Fermi bubble then it must produce $\sim10^{55}$\,erg; the corresponding efficiency is $\sim 6\ee -5 $, indicating that the efficiency of this accretion mode must be  significantly higher than at present.

The feeding of molecular gas into the central 100\,pc of the Galaxy is thought to be ongoing and related to the presence of the Galactic bar (e.g. Morris \& Serabyn 1996).  Infall of molecular gas into the central few parsec of the Galaxy appears to be continual, and so partial capture of gas clouds by Sgr A* may be quite common.  The cloud capture event responsible for the formation of the stellar disk, and potentially the Fermi bubble, occurred within the last few million years and added 1\% to the mass of Sgr A* and a similar mass to the local stellar population; such events may have contributed significantly to the growth of Sgr A* and the surrounding stellar population over the life time of the Galaxy.

This scenario may also apply to gas accretion by SMBH in external galaxies, where it has been argued that the difficulties in fuelling AGN through an extended accretion disk imply that black holes are fed by a series of small gas capture events that create a sub-0.1\,pc accretion disk (Goodman 2003; King \& Pringle 2007).  The partial cloud capture model predicts a quadratic dependence of the disk mass on black hole mass (WY12).  Recent studies suggest a correlation between star formation rate and average black hole accretion rate in star forming galaxies (e.g., Chen et al. 2013). In addition, there is evidence for $\gamma$-ray emission from starburst galaxies such as M82 and NGC 253 (Abdo et al 2010). This correlation and the energetic activity associated with starburst galaxies can be understood as arising through simultaneous star formation and accretion during a series of cloud capture events over the lifetime of the central black hole.

\acknowledgments 
We thank Yoram Lithwick for insightful comments.  This work is partially supported by grants DP0986386 and DP120101792 from the Australian Research Council and AST-0807400 from the NSF.
{}

\vfill\eject

\begin{figure}
\center
\includegraphics[scale=1.2,angle=0]{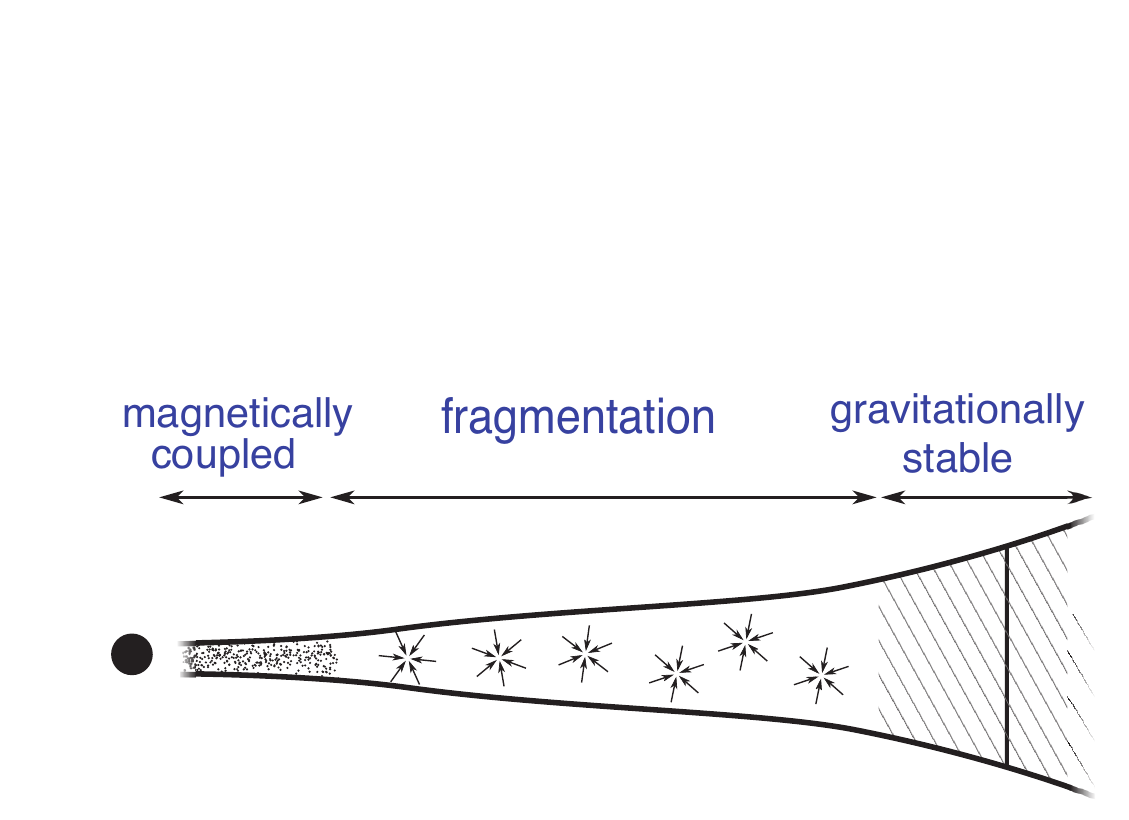}\\
\caption{
Schematic diagram of a gaseous disk orbiting Sgr A* (black dot).  If the surface density profile is steeper than $\sim r^{-1.5}$, self gravity is unimportant in the outer radii.  Fragmentation occurs at intermediate radii; but is suppressed by inefficient cooling at the innermost radii where the disk becomes very optically thick and magnetic activity drives accretion (see text).
}
\end{figure}  

\begin{figure}
\center
\includegraphics[scale=1.2,angle=0]{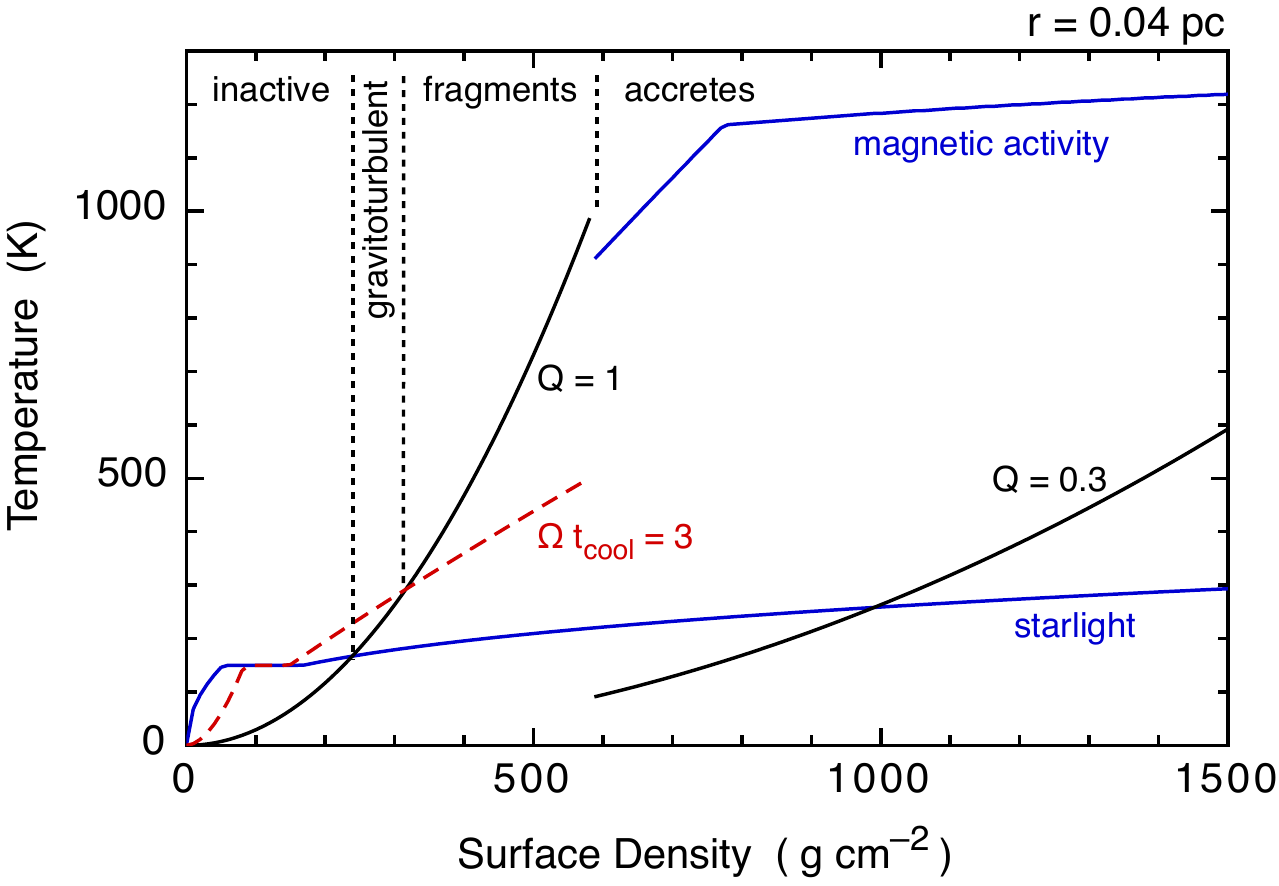}\\
\caption{
Dependence of characteristic midplane temperatures on disk surface density at 0.04 pc from Sgr A*.  Blue curves show the equilibrium temperature maintained by magnetic heating or irradiation by hot stars.  Black curves indicate the temperature below which the disk would be self-gravitating.   Red curve shows the temperature at which the disk's cooling time scale is $3\,\Omega^{-1}$.  Vertical black dotted lines indicate the column density ranges over which the disk is either inactive, or gravitoturbulent, or fragments, or accretes due to magnetic activity (see text).
}
\end{figure}  

\begin{figure}
\center
\includegraphics[scale=1.3,angle=0]{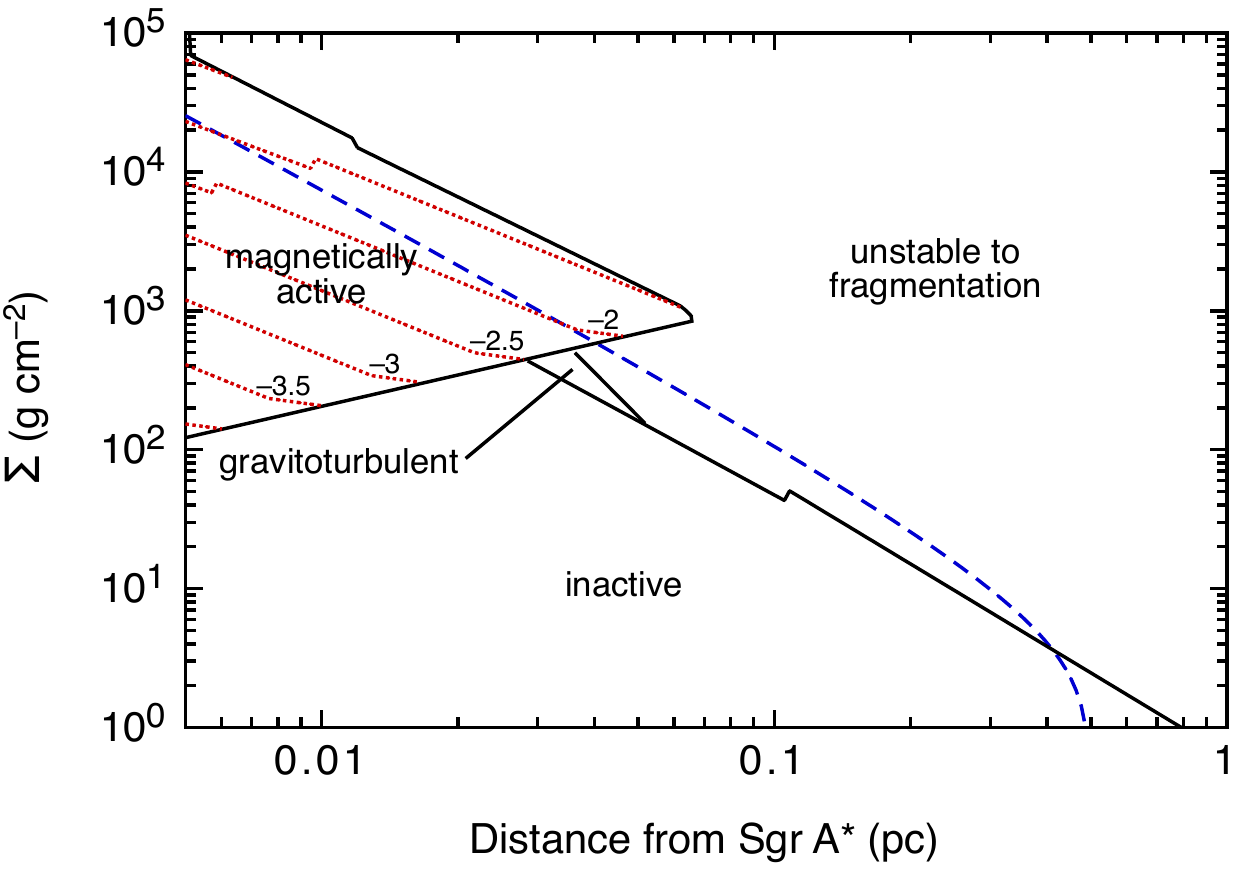}\\
\caption{
The dependence of the  fate of a localized region of a disk on distance from Sgr A* and surface density. At each radius, black lines delimit the range of surface densities for which the disk is inactive, or fragments, or settles into an accretion disk with turbulence driven either by the magnetic field or by self-gravity (see text).  The accretion rate is indicated by the red dotted contours,  labelled by $\log_{10}(\dot{M}/\msol\,$yr$^{-1})$.  The blue dashed curve indicates the surface density profile predicted by the capture model, for a total disk mass of $2\times10^5\,\msol$
}
\end{figure}  

\begin{figure}
\center
\includegraphics[scale=1.1,angle=0]{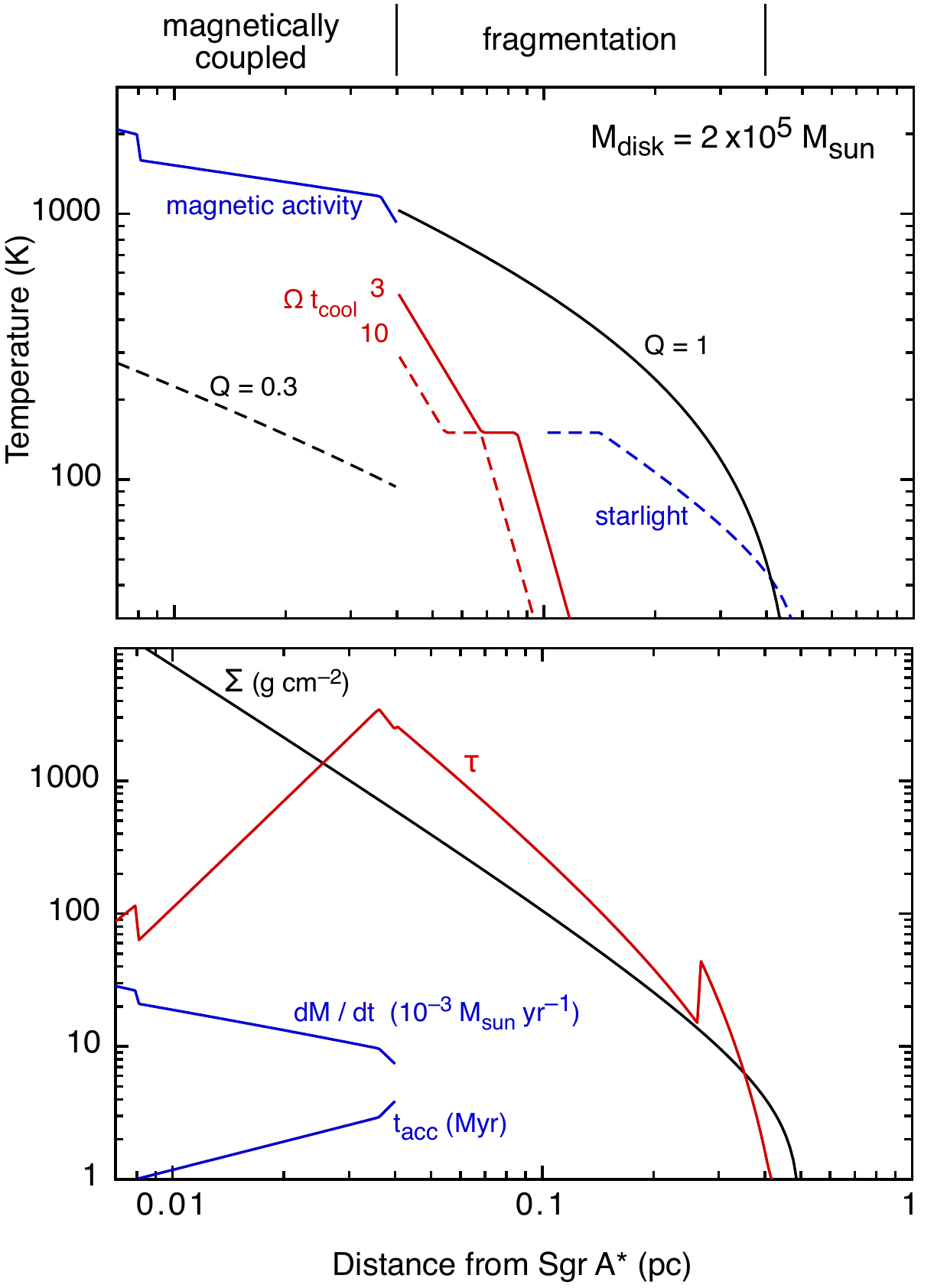}\\
\caption{
\emph{Upper panel:} Radial variation of the quantities in Fig.\ 2 for a disk mass of $2\times10^5\msol$, corresponding to $\Sigma \approx 600$\,g\,cm$^{-2}$ at 0.04\,pc from Sgr A*.  The disk fragments outside of 0.04\,pc.  Inside this radius it settles into a magnetically-mediated accretion disk (see text).  \emph{Lower panel:} Black and red curves show the adopted surface density profile and optical depth, respectively.   Blue curves show the local accretion rate and time scale.
}
\end{figure}

\end{document}